\documentclass[lettersize,journal]{IEEEtran}
\usepackage{amsmath,amsfonts}
\usepackage{algorithmic}
\usepackage{algorithm}
\usepackage{array}
\usepackage[caption=false,font=normalsize,labelfont=sf,textfont=sf]{subfig}
\usepackage{textcomp}
\usepackage{stfloats}
\usepackage{url}
\usepackage{verbatim}
\usepackage{graphicx}
\usepackage{cite}
\usepackage{color}
\begin{document}

\title{Energy Detection over Composite $\kappa-\mu$ Shadowed Fading Channels with Inverse Gaussian Distribution in Ultra mMTC Networks}

\author{He~Huang,~\IEEEmembership{Member,~IEEE,}
        Zeping~Sui,~\IEEEmembership{Member,~IEEE,}
        Zilong~Liu,~\IEEEmembership{Senior Member,~IEEE,}
        Wei~Huang,~\IEEEmembership{Senior Member,~IEEE,}
        Md.~Noor-A-Rahim,~\IEEEmembership{}
        Haishi Wang,~\IEEEmembership{Member,~IEEE,}
        Zhiheng Hu~\IEEEmembership{}
\thanks{He Huang, Haishi Wang and Zhiheng Hu are with College of Communication Engineering (College of Microelectronics), Chengdu University of Information Technology, Chengdu 610225, China. He Huang is also with Intelligent Interconnected Systems Laboratory of Anhui Province, Hefei University of Technology, Hefei 230009, China, e-mail: (huanghe@cuit.edu.cn; whs@cuit.edu.cn; hzh@cuit.edu.cn)}
\thanks{Zeping Sui and Zilong Liu are with the School of Computer Science and Electronics Engineering, University of Essex, Colchester CO4 3SQ, U.K. (e-mail: zepingsui@outlook.com, zilong.liu@essex.ac.uk).}
\thanks{Wei Huang is with School of Computer Science and Information Engineering, also with Intelligent Interconnected Systems Laboratory of Anhui Province, Hefei University of Technology, Hefei 230009, China, e-mail: (huangwei@nfut.edu.cn)}
\thanks{Md. Noor-A-Rahim is with nasc Research, School of Computer Science \& IT, University College Cork, Cork T12 XF62, Ireland, e-mail: (m.rahim@cs.ucc.ie)}
}

\IEEEpubid{}

\maketitle

\begin{abstract}
This paper investigates the characteristics of energy detection (ED) over composite $\kappa$–$\mu$ shadowed fading channels in ultra machine-type communication (mMTC) networks. We have derived the closed-form expressions of the probability density function (PDF) of signal-to-noise ratio (SNR) based on the Inverse Gaussian (\emph{IG}) distribution. By adopting novel integration and mathematical transformation techniques, we derive a truncation-based closed-form expression for the average detection probability for the first time. It can be observed from our simulations that the number of propagation paths has a more pronounced effect on average detection probability compared to average SNR, which is in contrast to earlier studies that focus on device-to-device networks. It suggests that for 6G mMTC network design, we should consider enhancing transmitter-receiver placement and antenna alignment strategies, rather than relying solely on increasing the device-to-device average SNR.
\end{abstract}


\section{Introduction}
During the last decade, spectrum efficiency has been considered an important performance metric of next-generation
wireless networks [1]-[4]. As a non-coherent signal processing
technology, energy detection (ED) has been widely used in
multiple scenarios, such as radar communication systems, ultra-wideband (UWB) systems, and millimeter-wave (mm-wave) communication networks [3], [4]. The authors of [4] initially studied the detection of unknown signals over flat band-limited Gaussian noise channels, focusing on a binary hypothesis-testing problem to derive the expressions for the probability of detection $P_d$ and the likelihood of false alarm $P_f$, which follow non-central and central chi-square distributions, respectively. Based on the favorable characteristics of ED, i.e., non-coherent structure and low implementation complexity, Digham \emph{et al.} examined its performance over classical fading channels such as Rayleigh, Rician, and Nakagami-\emph{m}, particularly in the context of radar communication systems, UWB communications, and mm-wave networks [3], [4].

As a parallel development, the generalized $\kappa$–$\mu$ distribution-based fading channel models were proposed to provide an accurate
small-scale characterization for line-of-sight (LoS) propagation. By selecting appropriate values for $\kappa $ and $\mu $
It encompasses several classical fading models as special cases, including Nakagami-\emph{m}, Rayleigh, Rician, and one-sided Gaussian
fading [5]-[7]. To characterize severe fading scenarios in mobile
radio propagation, the $\kappa$–$\mu$ extreme distribution was introduced in [8]. Furthermore, to incorporate both multipath and shadowing effects, composite shadowed fading models have also been explored for energy detection, including Nakagami-\emph{m}/lognormal, $\kappa$–$\mu$/lognormal, and $\eta$–$\mu$/lognormal fading channels [8]-[10]. In addition, the \emph{IG} distribution has been adopted to model the shadowing component, as it effectively captures the tail behavior of lognormal distributions with significant variance and enhances the probability of low-amplitude channel realizations, which is relative to the Gamma distribution [8]-[12]. However, the closed-form expression of the detection probability over LoS/lognormal fading channels is intractable to derive since integral functions are difficult to resolve.

Against this backdrop, we investigate energy detection over composite $\kappa  - \mu $ shadowed fading channels in ultra-dense mMTC networks in this paper. Our main contributions are summarized as follows.
\begin{itemize}
    \item With the aid of \emph{IG} distribution, we have transformed LoS/lognormal distribution into LoS/\emph{IG} distribution. Then, we derive the closed-form expressions of PDFs of envelope and instantaneous SNR for composite $\kappa  - \mu $ shadowed fading channels.
    \item To the best of our knowledge, this is the first investigation of deriving the new expression of average detection probability over composite $\kappa  - \mu $ shadowed fading channels, by leveraging a set of novel integration and mathematical transformation techniques. Besides, we introduce truncation-based approximations for practical evaluation. Furthermore, we have performed simulations to demonstrate that a finite set of truncated terms is sufficient to accurately evaluate the average detection probability.
    \item Through Monte Carlo simulations, we present a new insight into transmission behavior under composite $\kappa  - \mu $ shadowed fading channels: the transmission mode has a greater influence on the detection performance than the average SNR. By contrast, increasing the device-to-device SNR is enough to improve detection probability under conventional LoS fading scenarios [3], [4]. Our simulation results show that in composite $\kappa  - \mu $ shadowed fading channels, the number of propagation paths has a more significant impact on the detection probability than the average SNR. Therefore, in ultra-mMTC mobile networks, we shall jointly consider the number of paths and the average SNR to ensure reliable signal detection.
\end{itemize}

This paper is organized as follows: Section \uppercase\expandafter{\romannumeral2} introduces the ED system model and composite $\kappa  - \mu $ shadowed fading channels. In Section \uppercase\expandafter{\romannumeral3}, we derive the closed-form expression of average detection probability using ED. Section \uppercase\expandafter{\romannumeral4} illustrates our simulation results. Finally, we conclude this paper in Section \uppercase\expandafter{\romannumeral5}.

\section{ED Model and $\kappa  - \mu$ Fading with \emph{IG} Distribution}
The ED pattern is assumed to be a binary hypothesis-testing problem to determine the absence or presence of an unknown signal, which can be formulated as [6]

\begin{equation}
\begin{aligned}
\left\{ \begin{array}{l}
{H_0}:y(t) = n(t)\\
{H_1}:y(t) = hs(t) + n(t),
\end{array} \right.
\end{aligned}
\end{equation}
where $H_0$ denotes signal is absent while $H_1$ represents signal is present, $y(t)$ is the received signal, $n(t)$ denotes zero-mean complex additive white Gaussian noise (AWGN), $h$ is the wireless channel gain, and $s(t)$ denotes the transmitted signal. As illustrated in Fig. 1,  the received signal test statistics can be expressed as [6]:

\begin{equation}
\begin{aligned}
Y \sim \left\{ \begin{array}{l}
{H_0}:\chi _{2u}^2      \\
{H_1}:\chi _{2u}^2(2\gamma ),
\end{array} \right.
\end{aligned}
\end{equation}
where $u = TW$ denotes time bandwidth product with time interval $T$  and single-sided signal bandwidth $W$, $\chi _{2u}^2$ is a central chi-square distribution with degrees of freedom 2$u$, $\chi _{2u}^2(2\gamma )$ is a non-central chi-square distribution with degrees of freedom 2$u$ and a non-centrality parameter 2$\gamma $, where $\gamma $ denotes instantaneous SNR.
\begin{figure}[htbp]
  \centering
  \includegraphics[width=3.4 in]{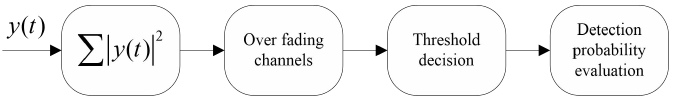}
  \caption{ED over composite $\kappa  - \mu $ shadowed fading channels.}
  \label{Fig.1}
\end{figure}
Consequently, the detection probability $P_d$ can be formulated as [6]
\begin{equation}
\begin{aligned}
{P_d} = {P_r}(y > \lambda |{H_1}) = {Q_u^{ED}}(\sqrt {2\gamma } ,\sqrt \lambda  ),
\end{aligned}
\end{equation}
where ${P_r}(\cdot)$ is probability, ${Q_u^{ED}}( \cdot , \cdot )$ is the \emph{u}-th order generalized Marcum ${Q}$-function, and $\lambda $ denotes the ED threshold.

The $\kappa  - \mu $ distribution is a generic fading distribution which illustrates small-scale and LoS fading scenario, which involves several clusters. Each cluster has multipath components with identical powers, and a dominant component has arbitrary power. The parameter $\kappa $ is the ratio between the total power of the dominant components and the total power of the scattered waves, and the parameter $\mu $ represents the related variable of multipath clusters. The envelope PDF can be expressed as [8]

\begin{equation}
\begin{aligned}
f(\rho ) =&\frac{{2\mu {{(1 + \kappa )}^{\frac{{\mu  + 1}}{2}}}}}{{{\kappa ^{\frac{{\mu  - 1}}{2}}}\exp (\mu \kappa )}}{\rho ^\mu }\exp [ - \mu (1 + \kappa ){\rho ^2}]\\
&{I_{\mu  - 1}}[2\mu \sqrt {\kappa (1 + \kappa )} \rho ],
\end{aligned}
\end{equation}
where $\rho $ is independent variable and ${I_v}(\cdot)$ denotes the modified Bessel function of the first kind. Based on the principles of composite shadowed statistical distribution, the PDF of composite $\kappa  - \mu$ shadowed channels can be derived as
\begin{equation}
\begin{aligned}
f_{\log }^{\kappa  - \mu }(r) = \int_0^\infty  {f_{R{\rm{|}}y}^{\kappa  - \mu }(r} {\rm{|}}y){f_{\log }}(y)dy,
\end{aligned}
\end{equation}
where $r$ is independent variable, $y = {\bar r^2}$ with $\bar r = \sqrt {\mathbb{E}({r^2})}$ where $\mathbb{E}(\cdot)$ is the expectation operator. Moreover, $f_{r{\rm{|}}y}^{\kappa  - \mu }(r|y)$ is the conditional envelope PDF of $\kappa  - \mu $ fading channels, $r$ denotes the envelop of fading signal and ${f_{\log }}(y)$ is the PDF of lognormal shadowing distribution. The conditional PDF of the $\kappa  - \mu $ distribution can be shown as [9]
\begin{equation}
\begin{aligned}
f_{R{\rm{|}}y}^{\kappa  - \mu }(r{\rm{|}}y) = &\frac{{2\mu {{(1 + \kappa )}^{\frac{{\mu  + 1}}{2}}}}}{{{\kappa ^{\frac{{\mu  - 1}}{2}}}\exp (\mu \kappa )}}\frac{{{r^\mu }}}{{{y^{\frac{{\mu  + 1}}{2}}}}}\exp \left[ { - \mu (1 + \kappa )\frac{{{r^2}}}{y}} \right]\\
&{I_{\mu  - 1}}\left[ {2\mu \sqrt {\kappa (1 + \kappa )} \frac{r}{{{y^{\frac{1}{2}}}}}} \right].
\end{aligned}
\end{equation}
Besides, the PDF of the lognormal shadowing distribution is given by [9], [10]
\begin{equation}
\begin{aligned}
{f_{\log }}(y) = \frac{\xi }{{\sqrt {2\pi } \sigma y}}\exp \left[ { - \frac{{{{(10\lg y - \psi )}^2}}}{{2{\sigma ^2}}}} \right],
\end{aligned}
\end{equation}
where $y$ is independent variable with the constant $\xi=4.342 $, $\psi $ and $\sigma $ are the mean and standard deviation of lognormal random variable $\lg y$. To obtain a more tractable expression to calculate (6), we can exploit the \emph{IG} distribution to present the lognormal shadowing distribution, yielding
\begin{equation}
\begin{aligned}
\eta  = \frac{{\exp (\psi )}}{{2\sinh ({\sigma ^2}/2)}},\theta  = \exp (\psi  + {\sigma ^2}/2),
\end{aligned}
\end{equation}
where $\eta$, $\theta $ are functions that include $\psi$, ${\sigma ^2}$. We define the polynomial as
\begin{equation}
\begin{aligned}
A{\rm{ = }}\frac{{\sqrt {2\eta } \mu {{(1 + \kappa )}^{\frac{{\mu  + 1}}{2}}}}}{{\sqrt \pi  {\kappa ^{\frac{{\mu  - 1}}{2}}}\exp (\mu \kappa
)}}.
\end{aligned}
\end{equation}
Consequently, the envelope PDF of composite $\kappa - \mu$ fading channels with $IG$ distribution is expressed as
\begin{equation}
\begin{aligned}
f_{\log /IG}^{\kappa  - \mu }(r) =& A\int_0^\infty  {{r^\mu }} \frac{1}{{{y^{\frac{\mu }{2} + 2}}}}{I_{\mu  - 1}}\left[ {2\mu \sqrt {\kappa (1 + \kappa )} \frac{r}{{{y^{\frac{1}{2}}}}}} \right]\\
&\exp \left[ { - \mu (1 + \kappa )\frac{{{r^2}}}{y} - \frac{\eta }{{2{\theta ^2}}}\frac{{{{(y - \theta )}^2}}}{y}} \right]dy,
\end{aligned}
\end{equation}
Then, with the help of [8], the PDF based on instantaneous SNR can be derived as (11), which is shown at the top of the next page, where $\bar \gamma $ is the average instantaneous SNR.
\begin{figure*}
\begin{equation}
\begin{aligned}
f_{\log /IG}^{\kappa  - \mu }(\gamma ) = \frac{{A\exp (\frac{\eta }{\theta }){\theta ^{\frac{\mu }{2} - 1}}}}{2}\int_0^\infty  {\frac{{{\gamma ^{\frac{\mu }{2} - \frac{1}{2}}}}}{{{{\bar \gamma }^{\frac{\mu }{2} + \frac{1}{2}}}}}\frac{1}{{{y^{\frac{\mu }{2} + 2}}}}} {I_{\mu  - 1}}\left[ {2\mu \sqrt {\kappa (1 + \kappa )} \frac{{{\theta ^{\frac{1}{2}}}{\gamma ^{\frac{1}{2}}}}}{{{{\bar \gamma }^{\frac{1}{2}}}{y^{\frac{1}{2}}}}}} \right]\exp \left[ { - \mu (1 + \kappa )\frac{{\theta \gamma }}{{\bar \gamma y}} - \frac{{\eta y}}{{2{\theta ^2}}} - \frac{\eta }{{2y}}} \right]dy.
\end{aligned}
\end{equation}
\hrule
\end{figure*}

\begin{figure}[t]
  \centering
  \includegraphics[width=0.8\linewidth]{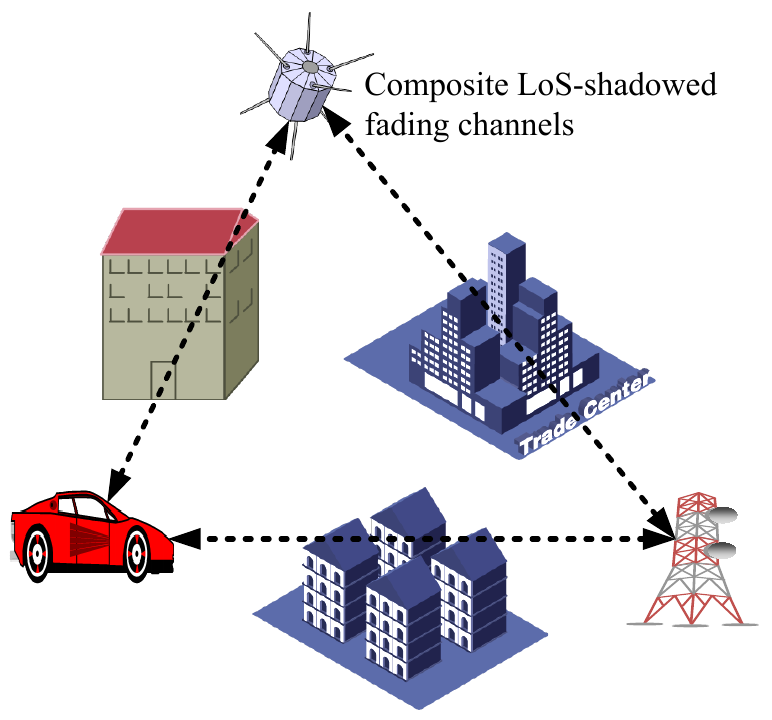}
  \caption{Composite $\kappa  - \mu $ shadowed fading channels in ultra-mMTC networks.}
  \label{Fig.1}
  \vspace{-1em}
\end{figure}
\section{Average Detection Probability over $\kappa  - \mu /IG$ Fading Model}

As shown in Fig. 2, in composite $\kappa  - \mu $ shadowed fading channels, the LoS channel components may be blocked due to physical obstructions. To capture this characteristic accurately, we investigate the detection probability of ED over composite $\kappa  - \mu$ fading channels with lognormal shadowing. The average detection probability of ED over the proposed $\kappa-\mu$/\emph{IG} fading model can be derived as

\begin{equation}
\begin{aligned}
\overline {{P_d}} _{\log /IG}^{\kappa  - \mu } = \int\limits_0^\infty  {Q_u^{ED}} (\sqrt {2\gamma } ,\sqrt \lambda  )f_{\log /IG}^{\kappa  -
\mu }(\gamma )d\gamma.
\end{aligned}
\end{equation}
The expression of the analytical detection probability can be formulated as [3]
\begin{equation}
\begin{aligned}
Q_u^{ED}(\sqrt {2\gamma } ,\sqrt \lambda  ) = \sum\limits_{l = 0}^\infty  {\frac{{\exp ( - \gamma ){\gamma ^l}\Gamma \left( {l + u,\frac{\lambda }{2}} \right)}}{{\Gamma (l + 1)\Gamma (l + u)}}} .
\end{aligned}
\end{equation}
where $\Gamma (\cdot)$ is the Gamma function and $\Gamma (\cdot,\cdot)$ is the incomplete Gamma function. We define the polynomial $B$ as
\begin{equation}
\begin{aligned}
B{\rm{ = }}A\frac{{\Gamma \left( {l + u,\frac{\lambda }{2}} \right)}}{{2\Gamma (l + 1)\Gamma (l + u)}}\exp \left( {\frac{\eta }{\theta }} \right){\theta ^{\frac{\mu }{2} - 1}}.
\end{aligned}
\end{equation}

Based on (11)-(14), the average detection probability can be derived as (15), which is shown
at the top of the next page. Since (15) involves integration variables $\gamma$ and $y$, we simplify the integral over $\gamma$ at first. We define the variable integral of the instantaneous SNR as
\begin{figure*}
\begin{equation}
\begin{aligned}
\overline {{P_d}} _{\log /IG}^{\kappa  - \mu } = \sum\limits_{l = 0}^\infty  {B\int\limits_0^\infty  {\int\limits_0^\infty  {\exp ( - \gamma )\frac{{{\gamma ^{l + \frac{\mu }{2} - \frac{1}{2}}}}}{{{{\bar \gamma }^{\frac{\mu }{2} + \frac{1}{2}}}}}{y^{ - \frac{\mu }{2} - 2}}} } } {I_{\mu  - 1}}\left[ {2\mu \sqrt {\frac{{\kappa (1 + \kappa )\theta \gamma }}{{\bar \gamma }}} \frac{1}{{{y^{\frac{1}{2}}}}}} \right]\exp \left[ { - \frac{{\eta y}}{{2{\theta ^2}}} - \frac{\eta }{{2y}} - \mu (1 + \kappa )\frac{{\theta \gamma }}{{\bar \gamma y}}} \right]d\gamma dy.
\end{aligned}
\end{equation}
\hrule
\end{figure*}
\begin{equation}
\begin{aligned}
{D_1}{\rm{ = }}&\int\limits_0^\infty  {{\gamma ^{l + \frac{\mu }{2} - \frac{1}{2}}}\exp \left[ { - \gamma  - \mu (1 + \kappa )\frac{{\theta \gamma }}{{y\bar \gamma }}} \right]} \\
&{I_{\mu  - 1}}\left[ {2\mu \sqrt {\frac{{\kappa (1 + \kappa )\theta \gamma }}{{\bar \gamma {y^{\frac{1}{2}}}}}} } \right]d\gamma.
\end{aligned}
\end{equation}
With the help of [13, (6.643.2)], we can simplify ${\gamma ^{\frac{1}{2}}}$ in ${I_v}(\cdot)$ and ${D_1}$ as
\begin{equation}
\begin{aligned}
{D_1} = &\frac{{\Gamma \left( {\frac{l}{2} + \mu } \right){y^{\frac{1}{2}}}{{\bar \gamma }^{\frac{1}{2}}}}}{{2\Gamma (\mu )\mu {\kappa ^{\frac{1}{2}}}{{(1 + \kappa )}^{\frac{1}{2}}}{\theta ^{\frac{1}{2}}}}}\exp \left[ {\frac{{{\mu ^2}\kappa (1 + \kappa )\theta }}{{2y\bar \gamma  + \mu (1 + \kappa )\theta }}} \right]\\
&{\left[ {\frac{{y\bar \gamma }}{{y\bar \gamma  + \mu (1 + \kappa )\theta }}} \right]^{l + \frac{\mu }{2}}}{M_{ - l - \frac{\mu }{2},\frac{{\mu  - 1}}{2}}}\left[ {\frac{{{\mu ^2}\kappa (1 + \kappa )\theta }}{{y\bar \gamma  + \mu (1 + \kappa )\theta }}} \right],
\end{aligned}
\end{equation}
where ${M_{u,v}}(z)$ denotes the Whittaker hypergeometric function. We define a polynomial $C$ as
\begin{equation}
\begin{aligned}
C{\rm{ = }}\frac{{\Gamma (l + \mu ){\mu ^{\mu  - 1}}{\kappa ^{\frac{{\mu  - 1}}{2}}}{{(1 + \kappa )}^{\frac{{\mu  - 1}}{2}}}{{\bar \gamma
}^l}}}{{\Gamma (\mu )}}.
\end{aligned}
\end{equation}
Upon leveraging [13, (9.220.2)] to simplify the exponential terms and the Whittaker hypergeometric function terms in (15) and (17), the average detection probability is derived as
\begin{equation}
\begin{aligned}
\overline {{P_d}} _{\log /IG}^{\kappa  - \mu } = &\sum\limits_{l = 0}^\infty  {BC} \int\limits_0^\infty  {\frac{{{y^{l - \frac{3}{2}}}\exp \left( { - \frac{{\eta y}}{{2{\theta ^2}}} - \frac{\eta }{{2y}}} \right)}}{{{{\left[ {\bar \gamma y + \mu (1 + \kappa )\theta } \right]}^{l + \mu }}}}}\\
&{}_1{F_1}\left[ {l + \mu ;\mu ;\frac{{{\mu ^2}\kappa (1 + \kappa )}}{{y\bar \gamma  + \mu (1 + \kappa )\theta }}} \right]dy,
\end{aligned}
\end{equation}
where ${}_1{F_1}(\cdot;\cdot;\cdot)$ is the confluent hypergeometric function. Therefore, we multiply the numerator and denominator of equation (19) by $y\bar \gamma$ and obtain
\begin{equation}
\begin{aligned}
\overline {{P_d}} _{\log /IG}^{\kappa  - \mu } = &\sum\limits_{l = 0}^\infty  {BC} \int\limits_0^\infty  {\frac{{{{(y\bar \gamma )}^{l - \frac{3}{2}}}\exp \left( { - \frac{{\eta y\bar \gamma }}{{2{\theta ^2}\bar \gamma }} - \frac{{\eta \bar \gamma }}{{2y\bar \gamma }}} \right)}}{{{{\bar \gamma }^{l - \frac{3}{2}}}{{\left[ {\bar \gamma y + \mu (1 + \kappa )\theta } \right]}^{l + \mu }}\bar \gamma }}}\\
&{}_1{F_1}\left[ {l + \mu ;\mu ;\frac{{{\mu ^2}\kappa (1 + \kappa )}}{{y\bar \gamma  + \mu (1 + \kappa )\theta }}} \right]d(y\bar \gamma ).
\end{aligned}
\end{equation}
In (20), we define the polynomial $D_2$ as
\begin{equation}
\begin{aligned}
{D_2} =& \int\limits_0^\infty  {\frac{{{{(y)}^{l - \frac{3}{2}}}\exp \left( { - \frac{{\eta y}}{{2{\theta ^2}\bar \gamma }} - \frac{{\eta \bar \gamma }}{{2y}}} \right)}}{{{{\left[ {y + \mu (1 + \kappa )\theta } \right]}^{l + \mu }}}}}\\
&{}_1{F_1}\left[ {l + \mu ;\mu ;\frac{{{\mu ^2}\kappa (1 + \kappa )}}{{y + \mu (1 + \kappa )\theta }}} \right]dy,
\end{aligned}
\end{equation}
where the confluent hypergeometric function $_1{F_1}(\cdot;\cdot;\cdot)$ [9] is derived as (22), which is shown at the top of the next page. We define $f = \mu (1 + \kappa)\theta$ and define polynomial $G(n)$ as
\begin{figure*}
\begin{equation}
\begin{aligned}
_1{F_1}\left[ {l + \mu ;\mu ;\frac{{{\mu ^2}\kappa (1 + \kappa )}}{{y + \mu (1 + \kappa )\theta }}} \right]{\rm{ = }}{\sum\limits_{n = 0}^\infty  {\frac{{{{(l + \mu )}_n}}}{{{{(\mu )}_n}}}\left[ {\frac{{{\mu ^2}\kappa (1 + \kappa )}}{{y + \mu (1 + \kappa )\theta }}} \right]} ^n}\frac{1}{{n!}}.
\end{aligned}
\end{equation}
\hrule
\end{figure*}

\begin{equation}
\begin{aligned}
G(n) = \frac{{{{(l + \mu )}_n}{\mu ^{2n}}{\kappa ^n}{{(1 + \kappa )}^n}}}{{{{(\mu )}_n}n!}}.
\end{aligned}
\end{equation}
where ${(\mu )_n} = \Gamma (\mu  + n)/\Gamma (\mu )$ is the Pochhammer symbol [13]. Upon expanding series expressions with infinite terms in (21) and letting $\xi=-l-\mu-n$, $D_2$ can be simplified as
\begin{equation}
\begin{aligned}
{D_2} =& \sum\limits_{n = 0}^\infty  {\int\limits_0^\infty  {G(n)} } {y^{l - \frac{3}{2}}}{(y + f)^{\xi}}\exp \left( { - \frac{{\eta y}}{{2{\theta ^2}\bar \gamma }} - \frac{{\eta \bar \gamma }}{{2y}}} \right)dy,
\end{aligned}
\end{equation}
Furthermore, we expand (24) based on Newton binomial series with a negative exponent [13] and obtain
\begin{equation}
\begin{aligned}
{D_2} = &\sum\limits_{n = 0}^\infty  {\sum\limits_{i = 0}^\infty  {G(n)C_\xi^i} } {[\mu (1 + \kappa )\theta ]^{\xi - i}}\\
&\int\limits_0^\infty  {{y^{l + i - \frac{3}{2}}}\exp \left( { - \frac{{\eta y}}{{2{\theta ^2}\bar \gamma }} - \frac{{\eta \bar \gamma }}{{2y}}} \right)} dy,
\end{aligned}
\end{equation}
where
\begin{equation}
\begin{aligned}
C_a^b= \frac{{a!}}{{(a - b)!b!}}.
\end{aligned}
\end{equation}
According to [13, (3.471.9)] and denote $\beta=l+i-\frac{1}{2}$, integral $D_2$ can be deduced as
\begin{equation}
\begin{aligned}
{D_2} =& 2\sum\limits_{n = 0}^\infty  {\sum\limits_{i = 0}^\infty  {G(n)C_\xi^i} } {\left[ {\mu (1 + \kappa )\theta } \right]^{\xi - i}}{(\theta \bar \gamma )^{\beta}}{K_{\beta}}\left( {\frac{\eta }{\theta }} \right),
\end{aligned}
\end{equation}
where ${K_v}(x)$ with ${K_{ - v}}(x) = {K_v}(x)$ is the \emph{v}-th order modified Bessel function of the second kind [13]. Finally, the detection probability over composite $\kappa  - \mu$ shadowed channels can be derived as
\begin{equation}
\begin{aligned}
\overline {{P_d}} _{\log /IG}^{\kappa  - \mu } = &\sum\limits_{l = 0}^{{l_0}} {\sum\limits_{n = 0}^{{N^*}} {\sum\limits_{i = 0}^\infty  {\frac{{\sqrt {2\eta } \Gamma \left( {l + u,\frac{\lambda }{2}} \right)\exp \left( {\frac{\eta }{\theta }} \right){\theta ^{\frac{\mu }{2} - 1}}}}{{\sqrt \pi  \exp (\kappa \mu )\Gamma (l + 1)\Gamma (l + u)\Gamma (\mu )}}} } }\\
&\Gamma (l + \mu ){\mu ^\mu }{(1 + \kappa )^\mu }{{\bar \gamma }^{\frac{1}{2}}}G(n)C_\xi^i{[\mu (1 + \kappa )\theta ]^{\xi - i}}\\
&{(\theta \bar \gamma )^{\beta}}{K_{\beta}}\left( {\frac{\eta }{\theta }} \right).
\end{aligned}
\end{equation}

\section{Simulation Results and Discussion}
This section presents simulation results for the proposed detection framework, focusing on the PDF of the instantaneous SNR and the average detection probability over composite $\kappa-\mu$ shadowed fading channels in device-to-device ultra-massive mMTC networks.

Fig. 3 illustrates the PDF of instantaneous SNR for the proposed $\kappa - \mu /IG$ fading model, with parameters of $\eta = 2$, $\theta = 1$, and average SNR $\bar{\gamma} = 10$ dB, for different values of $\kappa$ and $\mu$. It can be observed that increasing either $\kappa$ or $\mu$ results in sharper and more peaked PDF curves, indicating a higher probability density around the average SNR $\bar{\gamma} = 10$ dB. Notably, variations in $\mu$ have a more pronounced effect on the shape of the PDF curves compared to $\kappa$. This implies that the number of multipath clusters associated with $\mu$ can play a more dominant role in shaping the fading behavior than the ratio of the total power of dominant components to the scattered power, which corresponds to $\kappa$. Furthermore, the distribution becomes more concentrated around $\gamma = 10$ dB when we exploit the parameters $\kappa$ and $\mu$ with higher values, which is constence with the observations from other composite fading channels, involving $\alpha - \mu / \alpha - \mu$ and $\kappa - \mu / \alpha - \mu$ [7]-[12]. These results confirm that the proposed $\kappa-\mu/IG$ model effectively captures the characteristics and can be used to evaluate detection performance under multipath and shadowing conditions.

\begin{figure}[h]
  \centering
  \includegraphics[width=0.9\linewidth]{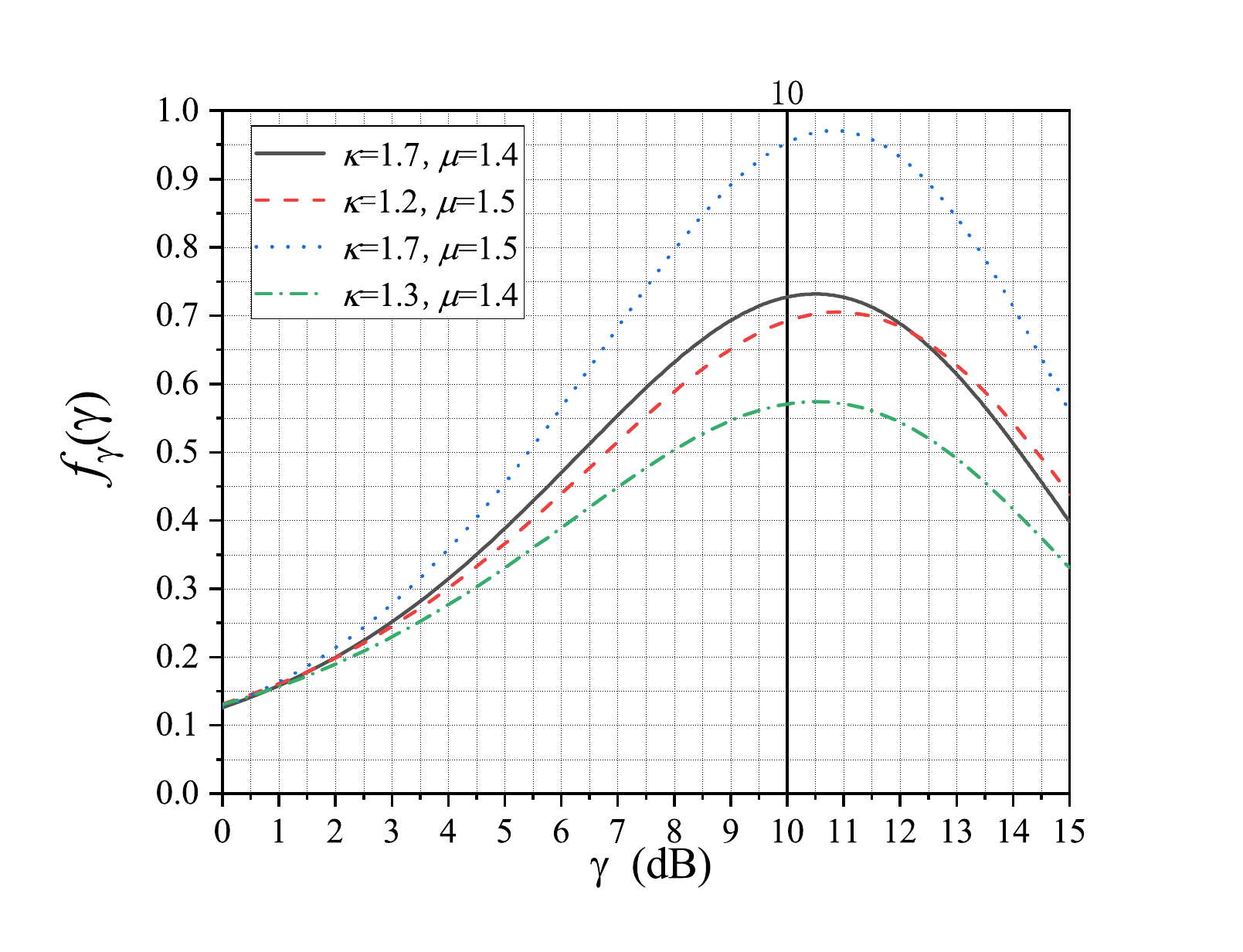}
  \caption{The PDF of SNR of $\kappa  - \mu /IG$ distribution using $\eta  = 2$, $\theta  = 1$ and $\bar \gamma  = 10$ with different $\kappa $ and $\mu $. }
  \label{Fig.1}
\end{figure}

To evaluate the approximate convergence behavior of the average detection probability, Fig. 4 presents simulations with varying numbers of truncation terms upon using $\kappa = 1.1$, $\mu = 1.2$, $\zeta = 2$, $\theta = 1$, $\lambda = 15$, and $u = 4$. Fig. 4 illustrates the relationship between the average detection probability and the average SNR under different truncation levels. It can be observed that, for fixed values of $\kappa$ and $\mu$, increasing the average SNR leads to improved detection performance. Moreover, the observations confirm that a suitable choice of truncation terms, as guided by the convergence condition in (28), yields accurate approximations with low computational complexity. When the number of truncation terms is increased beyond the minimum required for convergence, the detection probability estimates become increasingly precise. Moreover, it can be observed from Fig. 4 that there is a perfect overlap between the average detection probability closed-form expression and the Monte Carlo simulations, which validates the accuracy of our analytical results. The running time of calculating (28) is demonstrated in Table. I. Based on Fig. 4 and Table. I, we can see that there is a trade-off between complexity and average detection probability.

\begin{figure}[h]
  \centering
  \includegraphics[width=0.9\linewidth]{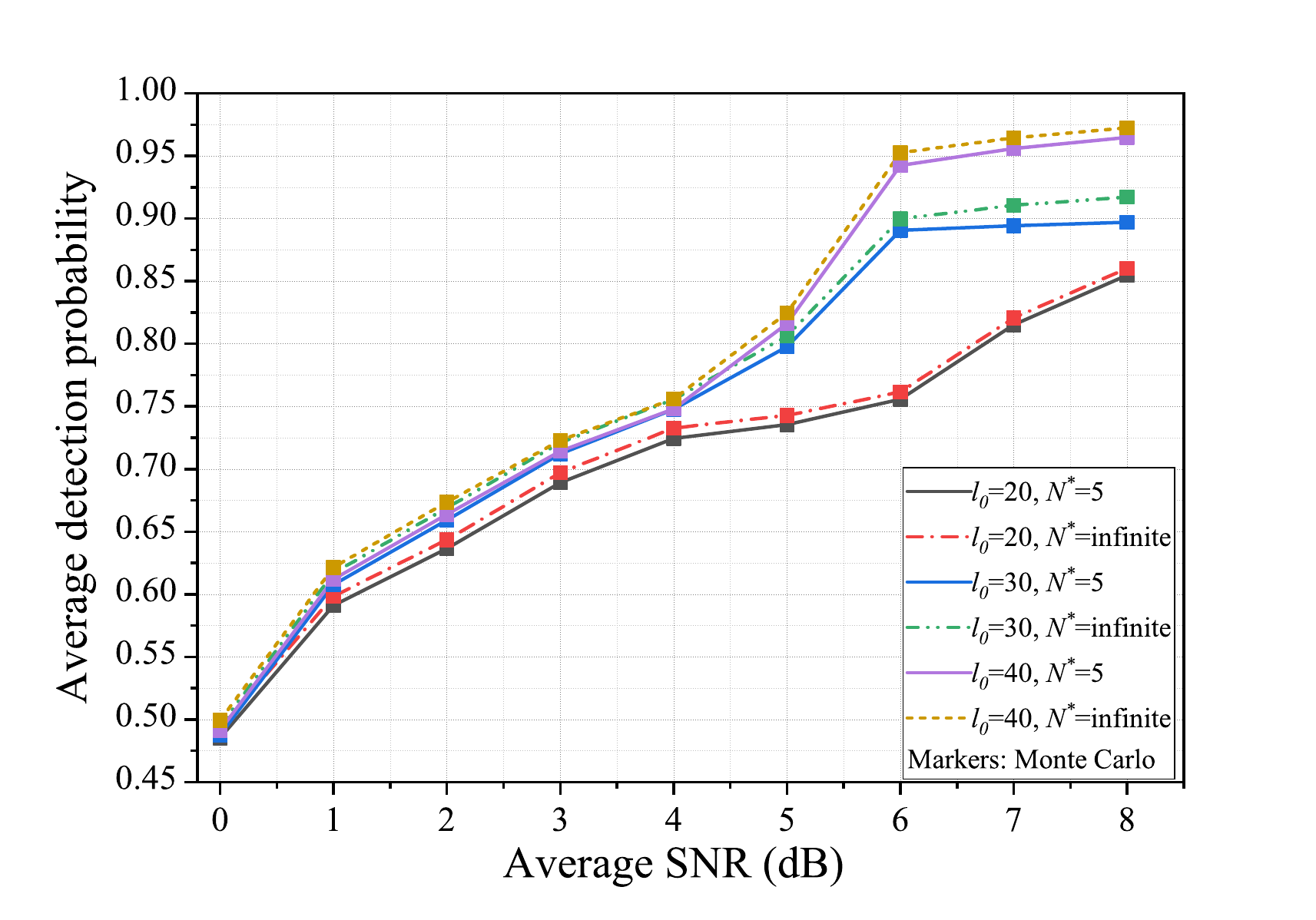}
  \caption{Average detection probability versus average SNR with truncation terms for $\kappa {\rm{ = }}1.1$, $\mu  = 1.2$, $\eta   = 2$, $\theta  = 1$ and $u = 4$.}
  \label{Fig.2}
\end{figure}

\begin{table}
  \centering
  \caption{Running time of the closed-form expression of (28) using the parameters in Fig. 4.}
  \begin{tabular}{l|c|c|c}
  \hline
Value of $l_0$ & $20$ & $30$ & $40$\\
    \hline
     Running time (s) & 1.0523 & 2.1142 & 3.1765\\
     \hline
  \end{tabular}
  \vspace{-2em}
\end{table}

In Fig. 5, we demonstrate that the proposed generalized fading model can be reduced to classical fading channels (e.g., Rayleigh or Rician) when specific values are assigned to $\kappa$ and $\mu$. Notably, the analysis also reveals that small increases in $\mu$ significantly enhance the average detection probability, whereas increasing the average SNR has a comparatively milder impact. This highlights the critical role of $\mu$, which represents the number of multipath clusters, in determining detection performance. Therefore, to effectively mitigate shadowing effects and enhance communication quality in 6G ultra-dense mMTC networks, priority should be given to increasing the number of multipath clusters. This can be achieved by optimizing the physical placement of transmitters and receivers or by adjusting the antenna orientation on mobile devices. Once favorable propagation conditions have been established (such as through optimal positioning of devices or antenna alignment), the transmit power can then be increased to enhance the average SNR and further improve the reliability of device-to-device communication.

\begin{figure}[h]
\vspace{-2em}
  \centering
  \includegraphics[width=0.9\linewidth]{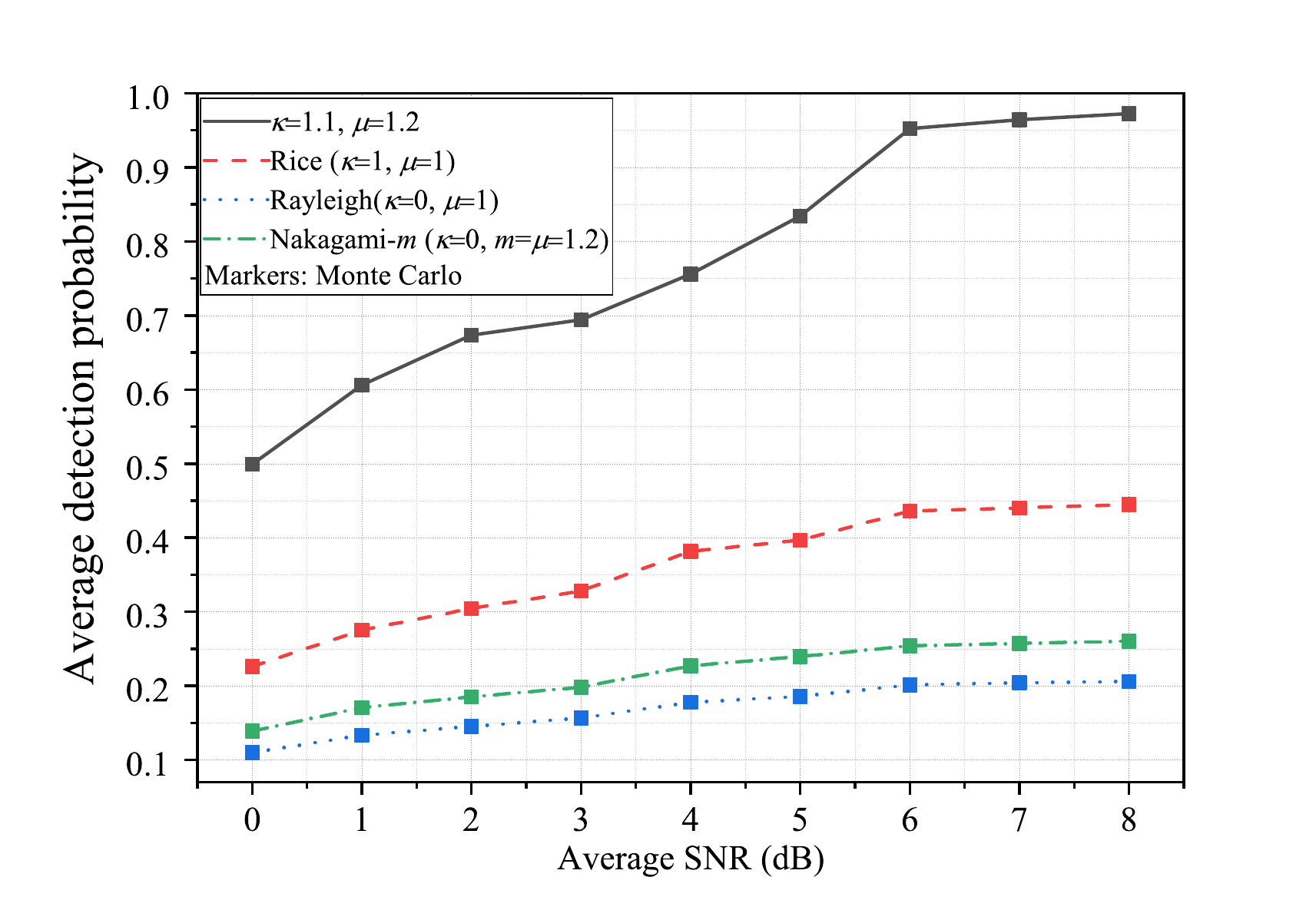}
  \caption{Average detection probability versus average SNR for classical fading models, ${l_0} = 40$, ${N^*} = {\text{infinite}}$, $\eta   = 2$, $\theta  = 1$ and $u = 4$.}
  \label{Fig.3}
\end{figure}

\section{Conclusion}
In this paper, we address the challenge of performing practical ED over composite $\kappa  - \mu $ shadowed fading channels in 6G ultra mMTC networks. We have derived the instantaneous PDF of the $\kappa$–$\mu$/lognormal fading distribution. Then, by using a series of new integration and transformation techniques, we have obtained a novel closed-form expression of the average detection probability for the first time. Furthermore, due to the difficulty of solving infinite series, we introduce a finite-term truncation approach and demonstrate that it provides an accurate estimation of the average detection probability. Importantly, our analysis reveals a key insight: in composite $\kappa-\mu$ shadowed fading environments, the number of multipath clusters has a more pronounced impact on detection performance than the device-to-device SNR. Based on this finding, we recommend that in 6G ultra-mMTC systems, communication strategies prioritize optimizing transmitter/receiver placement and antenna orientation to increase the number of multipath clusters before attempting to enhance the device-to-device SNR.


\bibliographystyle{ieeetr}

\end{document}